\documentclass{ws-mpla_arXiv12014160} 
\usepackage{graphicx}           

\newcommand{\beq}{\begin{equation}}
\newcommand{\eeq}{\end{equation}}
\newcommand{\beqa}{\begin{eqnarray}}
\newcommand{\eeqa}{\end{eqnarray}}
\newcommand{\bsubeqs}{\begin{subequations}} 
\newcommand{\esubeqs}{\end{subequations}} 
\begin{document}

\title{\vspace*{-1.1cm}Entropic-gravity derivation of MOND}

\author{F.R. Klinkhamer}
\address{Institute for Theoretical Physics, University of Karlsruhe,\\
             Karlsruhe Institute of Technology, 76128 Karlsruhe,
             Germany\\
             frans.klinkhamer@kit.edu}

\maketitle
\begin{abstract}
A heuristic entropic-gravity derivation has
previously been given of the gravitational two-body force
of modified Newtonian dynamics (MOND). Here, it is
shown that another characteristic of MOND can also be recovered,
namely, the external field effect (implying a violation of the
Strong Equivalence Principle).
In fact, the derivation gives  precisely the modified Poisson
equation which Bekenstein and Milgrom
proposed as a consistent nonrelativistic theory of MOND.
\vspace*{.5\baselineskip}\newline
Journal: \emph{Mod. Phys. Lett. A} \textbf{27}, 1250056 (2012).
\vspace*{.25\baselineskip}\newline
Preprint:  arXiv:1201.4160. 
\vspace*{.25\baselineskip}\newline
Keywords: Other theories of gravity,thermodynamics,
stellar dynamics and kinematics.
\vspace*{.25\baselineskip}\newline
PACS: 04.50.-h, 05.70.-a, 98.10.+z
\end{abstract}

\maketitle

\section{Introduction}\label{sec:Introduction}

One possible explanation of the missing-mass problem
on galactic scales is given by modified Newtonian dynamics
(MOND).\cite{Milgrom1983,BekensteinMilgrom1984,Milgrom1998}
An extensive and up-to-date review of the observational data
has been presented in Ref.~\refcite{FamaeyMcGaugh2011}.
Perhaps the \textit{experimentum crucis}
is provided by the nearby dwarf galaxy NGC 1560, where the
fine-structure in the rotation curve
is reproduced by MOND but not by a  `virialized'
dark-matter halo.\cite{Gentile-etal2010}
We thus have reason to investigate the theoretical basis of MOND.

MOND is, in fact, characterized by a new constant $a_0$ with the
dimensions of acceleration: considering the simplest case,
Newtonian dynamics is valid for accelerations $|\mathbf{a}|\gg a_0$
and modified dynamics for accelerations $|\mathbf{a}|\ll a_0$.
The question is what the fundamental origin of $a_0$ is.

A recent proposal starts from Verlinde's point of view
that the standard Newtonian gravitational force is really
an entropic force.\cite{Verlinde2010}
Assuming the microscopic degrees of freedom on the holographic screen
to have a nonzero minimum temperature $T_\text{min}$,
the MOND two-particle force between static test particles
can be derived,\cite{KlinkhamerKopp2011} together with
the result $a_0 \sim (c/\hbar)\, k_B T_\text{min}$ and an explicit
form of the interpolation function $\widehat{\mu}(x)$;
see below for details.
(A related paper on entropic gravity and MOND
is Ref.~\refcite{Pikhitsa2010}; additional references can be
found in Ref.~\refcite{KlinkhamerKopp2011}.
From a different point of view, several scaling relations
and numerical coincidences involving
MOND's acceleration scale $a_0$ have been discussed
in Ref.~\refcite{Bernal-etal2011}.)

A further characteristic of MOND is the so-called external field
effect (EFE).\cite{Milgrom1983} Loosely speaking,
the effect is that if a small weakly-bound gravitational
system (with internal acceleration $|\mathbf{a}_\text{int}| \ll a_0$)
is subjected to a large over-all acceleration
(with $|\mathbf{a}_\text{ext}| \gg a_0$), the internal gravitational dynamics
becomes standard Newtonian  even though
$|\mathbf{a}_\text{int}|/a_0$ is small.
This EFE corresponds to a violation of the
Strong Equivalence Principle, which suggests that the final
relativistic theory is fundamentally different from
general relativity.\cite{Will1993}

In this Letter,
we show that the entropic-gravity picture with $T_\text{min}>0$ also
explains the EFE. Moreover, we derive the
Bekenstein--Milgrom modified Poisson equation,\cite{BekensteinMilgrom1984}
which indeed provides a concrete realization of the EFE
and has other applications in the context of
missing-mass dynamics.\cite{FamaeyMcGaugh2011}

\section{Derivation}\label{sec:Derivation}

It is possible to closely follow Verlinde's
argument\cite{Verlinde2010} here.  The only new input is that,
for a general mass distribution, the nonspherical closed holographic
screen $\Sigma$ has a \emph{constant} minimum temperature
$T_\text{min}>0$ for the fundamental
degrees of freedom, in addition to the nonconstant temperature $T$
with $T\geq T_\text{min}$.
In fact, it is the difference $T-T_\text{min}$ which corresponds to
the effective localized mass that gravitates.
The derivation proceeds in five steps.

First, along with Verlinde (Sec. 3.4 and Fig.~4
of Ref.~\refcite{Verlinde2010}), interpret the holographic screen as an
equipotential surface, the quantity $-2\Phi/c^2$ being a measure of the
entropy per degree of freedom on the screen, having set $\Phi=0$ at
spatial infinity (where, loosely speaking, there is no emerged space).
Then, identify the acceleration of a test particle just outside
the screen as the opposite of the potential gradient,
\beq\label{eq:acceleration}
\mathbf{a} = - \boldsymbol{\nabla}\Phi \,.
\eeq
The basic idea is that inertia follows from the absence of an
entropy gradient, acceleration from the presence of an entropy gradient.
Specifically, the vector $\boldsymbol{\nabla}\Phi$,
if nonzero, is parallel to the outward normal unit-vector
$\boldsymbol{n}$ of the corresponding
surface element $\mathrm{d}A$ on the holographic screen,
$\boldsymbol{\nabla}\Phi \times \boldsymbol{n}=0$ and
$\boldsymbol{\nabla}\Phi \cdot \boldsymbol{n}>0\,$.

Second, recall the Davies--Unruh association between
acceleration and temperature (references can be found
in, e.g., Ref.~\refcite{Verlinde2010}).
According to Eq.~(6) of Ref.~\refcite{KlinkhamerKopp2011},
this gives\footnote{\label{ftn:dS-temp}The mathematical structure
of \eqref{eq:grad-phi-temperature} is the same as the one
from the Unruh-type
temperature\cite{Narnhofer-etal1996,DeserLevin1997}
of an idealized detector with constant linear acceleration
(for us, acceleration orthogonal to the screen)
in a de-Sitter universe with Hubble constant $H_\text{dS}$
(for us, $H_\text{dS} \equiv 2\pi k_B T_\text{min}/\hbar$).
As mentioned in the last two paragraphs of Sec.~3 in
Ref.~\refcite{KlinkhamerKopp2011} and referring to the
discussion in Refs.~\refcite{Narnhofer-etal1996,DeserLevin1997},
the choice of a de-Sitter spacetime as the
effective description of the minimum temperature of the fundamental
degrees of freedom is consistent with
the demands of local Lorentz invariance
(special relativity), which, surprisingly, will also be seen to play a role
in the third step of our derivation.}
\bsubeqs\label{eq:grad-phi-temperature-def-muhat-def-a0}
\beqa\label{eq:grad-phi-temperature}
\widehat{\mu}\left(\frac{|\mathbf{a}|}{a_0}\right)\;|\mathbf{a}|
&=&
2\pi\,(c/\hbar)\,\big[k_B T-k_B T_\text{min}\big]\,,
\eeqa\beqa   
\label{eq:def-muhat}
\widehat{\mu}(x) &\equiv& \sqrt{1+(2x)^{-2}} -(2x)^{-1} \,,
\eeqa\beqa  
\label{eq:def-a0}
a_0&\equiv& 4\pi\, (c/\hbar)\, k_B T_\text{min} \,.
\eeqa
\esubeqs
The functional behavior of \eqref{eq:def-muhat}
is such that $\widehat{\mu}(x) \to x$ for $x\to 0^{+}$
and $\widehat{\mu}(x) \to 1$ for $x\to\infty$.
Note that this particular function $\widehat{\mu}(x)$ has also
been suggested in Ref.~\refcite{Milgrom1998}, but without the link
between $\Delta T \equiv T-T_\text{min}$ and inertia or
gravity [see the discussion under Eq.~(9) in Ref.~\refcite{Milgrom1998}].
Precisely this link will appear in the next step of our derivation
[see also the discussion in the first paragraph
of Sec.~5 in Ref.~\refcite{KlinkhamerKopp2011}].

Third, replacing $T$ by $T-T_\text{min}$ in Eq.~(4.3) of
Ref.~\refcite{Verlinde2010} gives for the total active
(locally gravitating)  energy
\bsubeqs\label{eq:E-dA}
\beq\label{eq:E}
E\equiv M c^2 = \int_{\Sigma}\,\mathrm{d}N\,\frac{1}{2}\,
\big[k_B T-k_B T_\text{min}\big]
= \frac{1}{2\,f\, l^2}\,\int_{\Sigma}\,\mathrm{d}A\;
\big[k_B T-k_B T_\text{min}\big] \,,
\eeq
where we have used, in the first step, special relativity
and, in the last step, the following holographic relation
between the surface element $\mathrm{d}A$ and a subset $\mathrm{d}N$
of the total number $N$ of degrees of freedom on the screen:
\beq\label{eq:dA}
\mathrm{d}A= f\, l^2 \, \mathrm{d}N\,,
\eeq
\esubeqs
with $l^2$ the fundamental unit of area and $f$ a numerical
factor keeping track of possible internal degrees of
freedom of the `atom of area.'

Fourth, combining \eqref{eq:grad-phi-temperature} and \eqref{eq:E}
and recalling \eqref{eq:acceleration}, together with the
discussion below that equation, gives%
\bsubeqs\label{eq:M-GN}
\beq\label{eq:M}
M = \frac{1}{4\pi G_N}\,\int_{\Sigma}\,\;
    \widehat{\mu}\left(\frac{|\boldsymbol{\nabla}\Phi|}{a_0}\right)\;
    \boldsymbol{\nabla}\Phi \cdot \boldsymbol{n}\;\,\mathrm{d}A\,,
\eeq
with the identification
\beq\label{eq:GN}
G_N \equiv f\, c^3\,l^2/\hbar\,,
\eeq
\esubeqs
already encountered in our previous work
(quoted in Ref.~\refcite{KlinkhamerKopp2011}).

Fifth, consider a small mass $\widetilde{M}=\rho\,\widetilde{V}$
from a small volume $\widetilde{V}$
with surface $\widetilde{\Sigma}=\partial\widetilde{V}$
and approximately constant mass density $\rho$,
and bring this mass $\widetilde{M}$ close to the screen $\Sigma$.
According to Verlinde's argument given below Eq.~(4.4) in
Ref.~\refcite{Verlinde2010},
the following \emph{local} consistency condition holds for
pushing $\widetilde{M}$ into the screen:
\beq\label{eq:mod-Poisson}
\boldsymbol{\nabla} \cdot \left[\widehat{\mu}
\left(\frac{|\boldsymbol{\nabla}\Phi|}{a_0}\right)\,\boldsymbol{\nabla}\Phi\right]
 =4\pi G_N\;\rho\,.
\eeq
The underlying mathematics is trivial (but not the physics):
\eqref{eq:M} with $M$ replaced by
$\widetilde{M}$ and  $\Sigma$ by $\widetilde{\Sigma}$
(assuming the interior space with volume $\widetilde{V}$ to exist)
immediately gives \eqref{eq:mod-Poisson} by use of Gauss'
divergence  theorem and in the limit $\widetilde{V}\to 0$.

This completes our entropic-gravity derivation of the modified
Poisson
law \eqref{eq:mod-Poisson}.\footnote{\label{ftn:Pikhitsa}The
same equation has been obtained in Ref.~\refcite{Pikhitsa2010}
but with a different physics motivation.
Here, the starting point is the holographic screen with
its real physical degrees of
freedom,\cite{Verlinde2010,KlinkhamerKopp2011}
not an embedding spacetime with curvature.}
The derivation is, of course, entirely heuristic.
But, even without knowledge of the underlying microscopic theory,
the derivation may still be valid, as it relies on general principles:
thermodynamics and, in a more subtle way, special
relativity.\footnote{%
To a certain extent, inspiration is provided by Bohr's derivation
of the quantized energy levels of the atom
before the discovery of quantum theory proper.}

\section{Discussion}\label{sec:Discussion}

Equation \eqref{eq:mod-Poisson} corresponds, in fact,
to the particular modified Poisson law
suggested by Bekenstein and Milgrom,\cite{BekensteinMilgrom1984}
now with the explicit interpolation function
$\widehat{\mu}(x)$ from \eqref{eq:def-muhat}
and the MOND constant $a_0$ from \eqref{eq:def-a0}.

As shown by Bekenstein and Milgrom,\cite{BekensteinMilgrom1984}
the nonstandard gravitational field equation
\eqref{eq:mod-Poisson}  can be derived from a Lagrangian,
whose invariance under spacetime translations guarantees
energy and momentum conservation.
(For us, this Lagrangian is not a fundamental
object but an auxiliary quantity.)
Also noted by Bekenstein and Milgrom
(Sec.~V in Ref.~\refcite{BekensteinMilgrom1984})
is the external field effect (EFE) mentioned in the Introduction.
It is possible to illustrate the EFE from \eqref{eq:mod-Poisson}
by a simple one-dimensional example.\cite{FamaeyMcGaugh2011}

Consider an external mass distribution $\rho_\text{ext}$
which gives rise to the one-dimensional acceleration $g_\text{ext}$
and a local mass density $\rho$
which gives rise to the one-dimensional acceleration $g$.
The corresponding Newtonian accelerations
(from the standard Poisson law $\nabla^2\Phi=4\pi G_N\,\rho$)
are denoted $g_{N,\,\text{ext}}$ and $g_{N}$,
respectively. Using \eqref{eq:acceleration} and considering all vectors
to be one-dimensional (hence, no need for bold-face symbols),
it follows from \eqref{eq:mod-Poisson} that
\bsubeqs
\beq\label{eq:g+gext-tmp}
\nabla \cdot \left[
(g+g_\text{ext})\;\widehat{\mu}\big(|g+g_\text{ext}|/a_0\big)\right]
=
-4\pi G_N\;(\rho+\rho_\text{ext})\
=
\nabla \cdot \left(g_{N}+g_{N,\,\text{ext}}\right)\,,
\eeq
which gives
\beq\label{eq:g+gext}
(g+g_\text{ext})\;\widehat{\mu}\big(|g+g_\text{ext}|/a_0\big)
-g_\text{ext}\;\widehat{\mu}\big(|g_\text{ext}|/a_0\big)
=
g_{N} \,,
\eeq
\esubeqs
where the integration constant has been set to zero.
With the explicit function
\eqref{eq:def-muhat}, we obtain the following relation
between the local acceleration $g$ and the corresponding
Newtonian value $g_{N}$,
for given external acceleration $g_\text{ext}$:
\beqa\label{eq:g-general}
(g+g_\text{ext})\,
\left[\sqrt{1+(a_0/2)^2\,|g+g_\text{ext}|^{-2}}
-(a_0/2)\,|g+g_\text{ext}|^{-1}
\,\right]
&&
\nonumber\\[2mm]
-(g_\text{ext})\,
\left[\sqrt{1+(a_0/2)^2\,|g_\text{ext}|^{-2}}
-(a_0/2)\,|g_\text{ext}|^{-1}
\,\right]
&=&
g_{N}  \,,
\label{eq:roots}
\eeqa
which reduces to $g=g_{N}$ for $a_0=0$
or $T_\text{min}=0$ at a fundamental level.

For $|g_\text{ext}|\gg a_0$ and $|g|\ll |g_\text{ext}|$,
relation \eqref{eq:g-general} gives
the standard Newtonian dynamics (up to a
small renormalization of Newton's constant $G_N$),
\bsubeqs
\beq\label{eq:g-Newtonian}
g
\sim \left[1+\frac{1}{8}\,\left(\frac{a_0}{g_\text{ext}}\right)^2\,\right]\;
g_{N} \,,
\eeq
\noindent
having kept only the leading $(g_\text{ext})^{-2}$ term.
Observe that \eqref{eq:g-Newtonian} holds, in particular, for
$|g|\ll a_0$ and does not display a fundamental change from
the Newtonian behavior, which is precisely the content of the
EFE. For $|g_\text{ext}|\ll |g|\ll a_0$, on the other hand,
relation \eqref{eq:g-general} does give the MOND-type behavior,
\beq\label{eq:g-MONDian}
g\;|g|/a_0 \sim g_{N} \,,
\eeq
\esubeqs
which, as mentioned in the Introduction, may provide a solution to
the missing-mass problem on galactic
scales.\cite{Milgrom1983,FamaeyMcGaugh2011,Gentile-etal2010}

In a nonrelativistic flat-spacetime context,
our heuristic explanation of the EFE
is that a very large external acceleration
of a small self-gravitating system corresponds to
a temperature of the fundamental degrees of freedom
very much larger than their inherent minimum temperature,
so that deviations from the standard Newtonian gravitational
behavior become negligible.
Only if the external acceleration essentially
vanishes and if the internal acceleration is small enough,
is the net temperature of the fundamental degrees of freedom
small enough to reveal the existence of an inherent
(`hard-wired') minimum temperature $T_\text{min}$,
with the corresponding
non-Newtonian gravitational effects characterized
by the MOND acceleration scale $a_0\propto k_B T_\text{min}$.

Note that the previous entropic-gravity
derivation\cite{KlinkhamerKopp2011}
of the MOND constant $a_0 \propto k_B T_\text{min}$
held for the special case of linear relative motion between the
two test particles. This has now been generalized to
arbitrary (nonrelativistic) motions, and the result is
given by \eqref{eq:def-a0}.
The connection of the local-dynamics constant $a_0$
to a genuinely cosmological quantity can be only qualitative,
as the derivation, up till now, has been nonrelativistic.
Still, this connection\cite{KlinkhamerKopp2011} appears suggestive:
$a_0 \sim 2\,c\, H_\text{dS}$, from \eqref{eq:def-a0}
and the identification of
$T_\text{min}$ as the Gibbons--Hawking temperature
$T_\text{GH}=\hbar\,H_\text{dS}/(2\pi k_B)$
of a de Sitter universe with Hubble constant $H_\text{dS}$
(see also the remarks in Ftn.~\ref{ftn:dS-temp}).

With the external field effect from \eqref{eq:mod-Poisson}
implying a violation of
the Strong Equivalence Principle,\cite{BekensteinMilgrom1984,Will1993}
it can be expected that
the entropic-gravity derivation of the final relativistic version of
MOND, if successful, will result in a theory significantly different
from general relativity.\footnote{\label{ftn:relativity}Repeating
the steps of Sec.~\ref{sec:Derivation}, Verlinde's
relativistic derivation (Secs. 5.1--5.2 of Ref.~\refcite{Verlinde2010})
can again be closely followed.
This leads to a new version of his Eq.~(5.9):
$M$ $=$ $(4\pi G_N)^{-1}\,\int_{\Sigma}\, \exp[\phi/c^2]$
$\widehat{\mu}\big(|\nabla\phi|/a_0\big)$
$\nabla\phi\cdot \mathrm{d}A$,
in terms of an effective redshift-related potential
$\phi\equiv (c^2/2)\log[-\xi\cdot\xi]$
from the timelike Killing four-vector $\xi^{b}$ of the static
spacetime considered (with signature $-\,+\,+\,+$).
It remains to be seen if this type of surface integral
can give rise to a volume integral
which matches the volume integral corresponding to the mass $M$.}

\end{document}